\documentclass[aps]{revtex4}
\usepackage{graphicx}
\usepackage{natbib}

\newcommand{\etal}{{et~al.}}
\def\aa{{\sl Astron.\ \&\ Astrophys. \ }}

\def\apj{{\sl Astrophys.\ J. \ }}
\def\apjl{{\sl Astrophys.\ J.\ Lett. \ }}
\def\apjs{{\sl Astrophys.\ J.\ Supp. \ }}
\def\araa{{\sl Ann.\ Rev.\ Astron.\ Astrophys. \ }}

\def\ijmpd{{\sl International\ J.\ Mod.\ Phys. \ D \ }}
\def\jcap{{\sl J.\ Cosm.\ Astroparticle\ Phys. \ }}
\def\lrr{{Liv.\ Rev. Rel. \ }}
\def\mnras{{\sl MNRAS \ }}
\def\n{{\sl Nature \ }}

\def\np{{\sl Nucl.\ Phys. \ }}
\def\plb{{\sl Phys.\ Lett.\ B \ }}

\def\pr{{\sl Phys.\ Rep. \ }}
\def\prd{{\sl Phys.\ Rev.\ D \ }}
\def\prl{{\sl Phys.\ Rev.\ Lett. \ }}

\def\ptps{{Prog.\ Theor.\ Phys.\ Suppl. \ }}
\def\rmp{{\sl Rev.\ Mod.\ Phys. \ }}
\def\rpp{{\sl Rev.\ Progr.\ Phys. \ }}

\newcommand{\lsim}{\,\lower2truept\hbox{${<\atop\hbox{\raise4truept\hbox{$\sim$}}}$}\,}
\newcommand{\gsim}{\,\lower2truept\hbox{${>\atop\hbox{\raise4truept\hbox{$\sim$}}}$}\,}

\def\etal{{\rm et al.$\,$}}

\newcommand{\be}{\begin{equation}}
\newcommand{\ee}{\end{equation}}
\newcommand{\bea}{\begin{eqnarray}}
\newcommand{\eea}{\end{eqnarray}}

\begin{document}

\title{Weak lensing in generalized gravity theories}

\author{Viviana Acquaviva, Carlo Baccigalupi, Francesca Perrotta}
\affiliation{SISSA/ISAS, Via Beirut 4, 34014 Trieste, Italy, \\
INFN, Sezione di Trieste, Via Valerio 2, 34127 Trieste, Italy} 

\begin{abstract}
We extend the theory of weak gravitational lensing to cosmologies 
with generalized gravity, described in the Lagrangian by a generic 
function depending on the Ricci scalar and a non-minimal coupled scalar 
field. \\
We work out the generalized 
Poisson equations relating the dynamics of the fluctuating components 
to the two gauge invariant scalar gravitational potentials, fixing the new 
contributions from the modified background expansion and fluctuations. \\
We show how the lensing equation gets modified by the cosmic expansion as 
well as by the presence of the anisotropic 
stress, which is non-null at the linear level both in scalar-tensor gravity 
and in theories where the gravitational Lagrangian term features a 
non-minimal dependence on the Ricci scalar. 
Starting from the geodesic deviation, we derive the generalized expressions 
for the shear tensor and projected lensing 
potential, encoding the spacetime variation of the effective gravitational 
constant and isolating the contribution of the anisotropic stress, 
which introduces a correction due to the spatial correlation between the 
gravitational potentials. \\
Finally, we work out the expressions of the lensing convergence power spectrum 
as well as the correlation between the lensing potential and the Integrated 
Sachs-Wolfe effect affecting Cosmic Microwave Background total intensity and 
polarization anisotropies. \\
To illustrate phenomenologically the new effects, we work out approximate
expressions for the quantities above in Extended Quintessence scenarios 
where the scalar field coupled to gravity plays the role of the dark energy. 
\end{abstract}

\maketitle

\section{Introduction}
\label{i}

In the recent years we reached a remarkable convergence 
in the most important cosmological parameters describing 
the cosmic content as well the statistics of perturbations. 
The Universe is nearly geometrically flat, with an expansion rate 
of about 70 km/sec/Mpc, and structures 
grown out of a primordial linear spectrum of nearly Gaussian and scale 
invariant perturbations in the distribution of the energy density. 
About 5\% of the critical energy density is made of baryons, while
the remaining dark part is supposed to interact at most weakly with
the baryons themselves, since we observe it only through its
gravitational effects. 
The dark component appears to be 30\% pressureless, like in Cold Dark Matter 
(CDM) scenarios, dominating the gravitational potentials perturbations which 
host visible structures like galaxies or clusters. 
The remaining 70\% should be in some sort of vacuum energy, with negative 
pressure acting as a repulsive gravity, and responsible for a late time cosmic 
acceleration era. The case for this ``concordance cosmological model''
is now quite robust, supported by several independent datasets: the
distant type Ia supernovae (hereafter SNIa \cite{riess_etal,perlmutter_etal}), 
the Cosmic Microwave Background (CMB) anisotropies (see \cite{bennett_etal} and
references therein), the Large Scale Structure 
(LSS, \cite{percival_etal,dodelson_etal}) 
and the Hubble Space Telescope (HST \cite{freedman_etal}). 

The picture is clearly far from being satisfactory: in particular, 
without a better insight into the nature of the dark cosmological 
component, we cannot claim to have a satisfactory physical 
understanding of cosmology. The simplest description of the vacuum 
energy responsible for cosmic acceleration is a purely geometric term 
in the Einstein equations, the Cosmological Constant. On the other
hand, while the CDM has a well established support from the theories 
beyond the standard model of particle physics, a Cosmological Constant 
providing the 70\% of the critical density today raises two problems.  
The first is the fine tuning required to fix the vacuum energy scale 
about 120 orders of magnitude less than the Planck energy density
which is supposed to be the unification scale of all forces in the
early Universe. The second is a coincidence issue, simply why among
all the small non-zero values of the Cosmological Constant, 
the value was chosen to be comparable to the critical energy density
today. These questions, still largely unsolved, could be answered only 
if the concept of Cosmological Constant is extended to a more general 
one, admitting a dynamics of the vacuum energy, known now as the dark energy 
(see \cite{sahni_starobinski,peebles_ratra,padmanabhan} and references therein). \\
The simplest generalization, already introduced well before the
evidence for cosmic acceleration \cite{wetterich,ratra_peebles}, is a scalar 
field, dynamical and fluctuating, with a background evolution slow enough to
mimic a constant vacuum energy given by its potential, providing 
cosmic acceleration. As soon as the latter was discovered, a renewed
interest in these models appeared immediately \cite{coble_etal,ferreira_joyce}. 
In particular, it was demonstrated how the dynamics of this component, 
under suitable potential shapes inspired by super-symmetry and 
super-gravity theories (see \cite{masiero_etal} and \cite{brax_martin}, 
respectively, and references therein), can possess attractors in the
trajectory space, named tracking solutions, capable to reach the
present dark energy density starting from a wide set of initial 
conditions in the very early universe, thus alleviating, at least
classically, the problem of fine-tuning \cite{liddle_scherrer,steinhardt_etal}. 
The scalar field playing the role of the dark energy was named
Quintessence. Its coherent insertion among the other cosmological 
components allowed to constrain it from the existing data 
\cite{doran_etal,corasaniti_copeland,balbi_etal,baccigalupi_balbi_etal,
caldwell_etal,spergel_etal,melchiorri_etal,wang_tegmark}, 
as well as to investigate the relation of the dark energy with the 
other cosmological components: the explicit 
coupling with baryons is severely constrained by observations
\cite{carroll}, while the possible coupling of the Quintessence with the
Ricci scalar
\cite{sahni,chiba,uzan,bartolo_pietroni,perrotta_baccigalupi_etal,
faraoni,baccigalupi_matarrese_etal,esposito-farese_polarski,riazuelo_uzan,torres,
perrotta_baccigalupi,perrotta_matarrese_etal,linder,matarrese_baccigalupi_etal}
and the dark matter
\cite{amendola_2000,amendola_etal,matarrese_pietroni_etal,amendola_2003}, 
as well as the phenomenology 
arising from generalized kinetic energy terms 
\cite{armendariz-picon_etal,caldwell,malquarti_etal} have been extensively 
studied. \\
The generalization of cosmology that we consider here concerns the 
gravitational sector of the fundamental Lagrangian, admitting a 
general dependence on the Ricci and Brans-Dicke scalar fields. 
This subject is interesting per se (see \cite{fujii_maeda,hwang} and
references therein), and is receiving more attention after the
discovery of cosmic acceleration, with the attempt to interpret the
evidence for dark energy as a manifestation of gravity; this scenario 
has been recently proved to have relevant consequences for what 
concerns the dark energy fine-tuning problem mentioned above 
\cite{matarrese_baccigalupi_etal}. 

The gravitational lensing effect in 
cosmology is gathering a great interest becoming one of 
the most promising tools 
to investigate cosmological structures 
(see \cite{bartelmann_schneider} and references therein), and commonly thought 
in terms of strong and weak regime. The first one concerns 
highly magnified sources, generically through the generation 
of multiple images of a background object by a single lens with 
the typical size of a galaxy. \\
The weak gravitational lensing, which is the subject of the present
work, produces weak amplification, generically through the distortion 
of the pattern of background light; moreover, it is generated by a 
large set of scales and objects, ranging roughly from non-linear 
structures like galaxy clusters to the large scale distribution of
matter, still in linear regime. The weak lensing shear was 
detected recently by independent groups with 
astonishing agreement \cite{bacon_etal,wilson_etal,wittmann_etal,
maoli_etal,vanwaerbecke_etal}. Although the precision of such 
measurements does not allow to constrain different cosmological 
models, the planned observations will become certainly 
a crucial tool to investigate the behavior of dark matter and energy 
during the structure formation process and at the onset of 
cosmic acceleration \cite{refregier}. \\
In particular, several authors considered the possibility to 
investigate the dark energy component through the weak lensing, 
with different approaches ranging from shear distortion of background 
galaxies from clusters of galaxies to the weak lensing power causing 
a non-Gaussian pattern into the CMB anisotropies \cite{huterer,hu_2002,
bartelmann_etal,
bernardeau,weinberg_kamionkowski,bartelmann_meneghetti_etal,jain_taylor,
giovi_etal}. 
The reason of this interest is the timing: the structure formation, and 
the weak lensing carrying its physical information, occurs at an epoch 
which overlaps with the onset of cosmic acceleration; by virtue of 
this fact, it is reasonable to expect a good sensitivity of the weak 
lensing effect to the main dark energy properties such as the equation 
of state and its redshift behavior. \\
These studies are entering in an higher level of sophistication: 
the first outcome of N-body simulations in several dark energy 
scenarios have been published recently \cite{klypin_etal,dolag_etal,
maccio_etal}, 
mainly studying the impact of the background rate of expansion on 
the internal parameter of structures. The implementation of a light 
ray tracing technique through those structures allows to check 
numerically the weak lensing pattern produced by mildly and full 
non-linear density perturbations \cite{maccio}. 

The theory of weak gravitational lensing in ordinary 
cosmologies, i.e. made by radiation, dark matter and cosmological constant, 
is known and well established, but a comprehensive treatment in 
cosmologies with generalized theories of gravity still lacks in literature. 
The aim of the present work is to fill this gap, by providing the community 
with the recipe to interpret the weak lensing observations in a more 
general context. 
We follow the harmonic approach to weak lensing \cite{hu} and the 
treatment for generalized cosmological scenarios including 
cosmological perturbation \cite{hwang}, already exploited for 
investigating the effects of the explicit coupling between dark 
energy and gravity (see \cite{matarrese_baccigalupi_etal} and 
references therein). 

This work is organized as follows. In Section II 
we describe the cosmological models we deal with, for 
background and linear perturbations. In Section 
III we write the generalized Poisson equations 
relating the fluctuations in metric and cosmological 
components. The derivation and discussion of the lensing 
equation and weak lensing potential are in Section IV and 
V, respectively. Finally, in Section VI we draw the 
conclusions. 

\section{Generalized cosmologies}
\label{gc}

We shall consider a class of theories of gravity whose action is 
written in natural units as
\be
\label{action}
S = \int d^4x \sqrt{- g}\left[ \frac{1}{2\kappa}
f(\phi,R) - \frac{1}{2} \omega(\phi)\phi^{;\mu}\phi_{;\mu} -
V(\phi) + {\cal{L}}_{\rm{fluid}}\right],
\ee
where $g$ is the determinant of the background metric,
$R$ is the Ricci scalar, $\omega$ generalizes the kinetic term,
and ${\cal{L}}_{\rm{fluid}}$ includes contributions from the matter and
radiation cosmological components; $\kappa=8\pi G_{*}$ plays the role
of the ``bare'' gravitational constant, \cite{esposito-farese_polarski}.
Here as throughout the paper Greek indices run from 0 to 3, Latin
indices from 1 to 3.\\
The usual gravity term $R/16\pi G$ has been generalized by the general 
function $f/2\kappa$ \cite{hwang,perrotta_baccigalupi_etal}. Note that 
the formalism adopted is suitable for describing non-scalar-tensor gravity 
theories, i.e. without $\phi$, where however the dependence on $R$ is 
generic. 

\subsection{Einstein equations}
\label{eq}

By defining $F(\phi,R)=(1/\kappa)\partial f/\partial R$, the Einstein
equations
$ G_{\mu\nu}=\kappa T_{\mu\nu} $
take the following form:
\bea
\label{emt}
G_{\mu\nu}=T_{\mu\nu}=
& \frac{1}{F}\left[ T^{\rm{fluid}}_{\mu\nu} +
  \omega\left(\phi_{,\mu}\phi_{,\nu} -
\frac{1}{2}g_{\mu\nu} \phi_{,\sigma} \phi^{,\sigma}\right) +
g_{\mu\nu}\frac{f/\kappa - R F - 2 V}{2} + F_{,\mu;\nu} - g_{\mu\nu}F^{;
\sigma}_{; \sigma}\right]\,,\nonumber 
\eea 
where again one can
recognize a part depending on the fluid variables, and a part
relative to the non-minimally coupled scalar field of the theory
plus a contribution arising from the generalized gravity coupling
represented by the function $1/F$; for practical purposes we will
render this splitting explicit rewriting ${ T}_{\mu\nu}$ as 
\be
T_{\mu\nu} = \frac{1}{F}T_{\mu\nu}^{\rm{fluid}} + 
{\cal T}^{\rm{gc}}_{\mu\nu}\ . 
\ee 
Note that in our scheme the generalized cosmology term is active also if 
gravity is the same as in general relativity, and a minimally coupled scalar 
field, like in the Quintessence models, represents the only new ingredient 
with respect to the ordinary case. This is true also in the limiting 
case where the scalar field reduces to a Cosmological Constant. 
Moreover, notice that 
${\cal T}^{\rm{gc}}_{\mu\nu}$ is not conserved if gravity differs from 
general relativity; nonetheless, 
contracted Bianchi identities still hold and ensure \be \label{cons}
(T^{\rm{fluid}}_{\mu\nu})^{; \mu} = 0\,, \ee which allows to
derive the equations of motion for the matter variables only,
leading to a remarkable simplification \cite{hwang}. The 
expression for the stress-energy tensor relative to the scalar
field which is conserved in the generalized scenarios described by 
the action (\ref{action}) must include the term accounting for the
interaction with the gravitational field
\cite{perrotta_baccigalupi}. It is also worth to notice how the
equations (\ref{emt}) get simplified if the function $f$ is a
product of $R$ times a function of the field only: \be \label{nmc}
\frac{f}{\kappa}=RF\ . \ee In the following, we will refer to this
class of cosmologies as Non-Minimally Coupled (NMC) models. \\
We willingly keep this work as much general as possible, due to the 
large variety of scenarios covered by the action \ref{action}. 
We shall only consider the Extended Quintessence 
(EQ \cite{perrotta_matarrese_etal,baccigalupi_matarrese_etal}) scenario 
as an example to illustrate the new aspects of the weak lensing process 
with respect to ordinary cosmologies; in that cases, the field $\phi$, 
non-minimally coupled to gravity, also represents the dark energy, 
providing acceleration through its potential $V$. 
Specifically, the original works considered a NMC 
model defined as 
\be
\label{EQ}
F(\phi )=\frac{1}{\kappa}+\xi\phi^{2}\ ,
\ee
and an inverse power law potential $V(\phi )=M^{4+\alpha}/\phi^{\alpha}$ 
providing cosmic acceleration today. The constraints from solar-system 
experiments force the correction to the gravitational constant to be small 
in this specific models. Therefore it is suitable to make approximations 
to illustrate a sort of first order variation of the weak lensing in generalized 
gravity theories with respect to the case of ordinary cosmology. 

\subsection{Background}

We will write the unperturbed Friedmann Robertson Walker (FRW) metric
in spherical coordinates as:
\begin{equation}
\label{FRW}
ds^2 = a^2(\tau)\left(-\, d\tau^2 + \frac{1}{1 - K r^2}dr^2 +
r^2 d\Omega\right)\,,
\ee
where $K$ is the uniform spatial curvature of a spherically symmetric
three-space, $d\Omega$ is the metric of the two-sphere,
and $\tau$ stands for the conformal time variable, related to cosmic
time by the usual relation $dt = a(\tau)\,d\tau$. \\
The energy-momentum tensor (\ref{emt}) can be recast in a
perfect-fluid form: \be T_{\mu\nu} = (p + \rho)u_{\mu}u_{\nu} +
p\,g_{\mu\nu}\,; \ee the corresponding background energy density
and pressure are easily computed to be: 
\be 
\label{generalizedrho}
\rho =
\frac{1}{F}\left(\rho_{\rm{fluid}} + \frac{\omega}{2
a^2}{\phi^{\prime}}^2 + \frac {R F - f/\kappa}{2} + V - \frac{3
{\cal H} F^{\prime}}{a^{2}}\right) =\frac{1}{F}\rho_{\rm{fluid}}+
\rho_{\rm{gc}}\, , 
\ee 
\be 
\label{generalizedp}
p = \frac{1}{F}\left(p_{\rm{fluid}} +
\frac{\omega}{2  a^2}{\phi^{\prime}}^2- \frac{R F - f/\kappa}{2} -
V + \frac{F^{\prime\prime}}{a^2} +
\frac{{\cal H}F^{\prime}}{a^2}\right)=
\frac{1}{F}p_{\rm{fluid}}+p_{\rm{gc}}\,; 
\ee 
the prime denotes
differentiation with respect to conformal time and ${\cal H}$ is
the conformal Hubble factor $a^{\prime}/a$. As above, 
$\rho_{\rm{gc}}$ and $p_{\rm{gc}}$ do not obey the conservation law in 
ordinary cosmologies, $\rho_{\rm{gc}}^{\prime}+3{\cal 
H}(\rho_{\rm{gc}}+p_{\rm{gc}})\ne 0$.\\
In FRW cosmologies the expansion equation reads 
\be
\label{friedmann} {\cal H}^{2}=a^{2}\rho -K\,, 
\ee 
and it cannot be solved directly due to the appearance of ${\cal H}$ 
in $\rho_{\rm{gc}}$, which is explicit in the last term but is also 
contained 
in $RF-f/\kappa$ through 
\be
\label{r}
R=\frac{6}{a^{2}}\left(\dot{\cal H}+{\cal H}^{2}\right)\ .
\ee
Note that this is true also in theories where $f\equiv f(R)$ 
and no scalar field is present. \\
On the other hand, NMC scenarios admit a formal solution, which is 
\be \label{friedmann_solved} 
{\cal H}=-\frac{3}{2}\frac{F^{\prime}}{F}+\sqrt{
\frac{9}{4}\left(\frac{F^{\prime}}{F}\right)^{2}+
\frac{1}{F}\left(a^{2}\rho_{\rm{fluid}}+
\frac{\omega}{2}\phi^{\prime\,2}+a^{2}V\right)-K}\ , 
\ee 
where we have selected the expansion solution with positive ${\cal H}$. 
Note that the dependence of the comoving distances $r$ on the
redshift $z=1/a -1$ gets also modified, according to
(\ref{friedmann_solved}): 
\be \label{rz}
r=\int_{0}^{z}\frac{dz}{H(z)}\,. 
\ee 
We generically indicate the
single components in the fluid with $x$. Since
$T_{\rm{fluid}}^{\mu\nu}$ is conserved, energy density, pressure,
equation of state and sound velocity, defined as 
\be \rho_{x} = -
T^{0}_{0\, x} \ ,\ p_{x} = 1/3 \: T^{i}_{i\, x} \ ,\ w_{x} 
= p_{x}/\rho_{x}\ , \
c_{s}^{2}=p^{\prime}_{x}/\rho^{\prime}_{x}\ , 
\ee 
give rise to
conservation equations having the familiar form: 
\be
\label{conservation_background} \rho_{x}^{\prime}+3{\cal
H}\rho_{x}(1+w_{x})=0\ . 
\ee 
The last ingredient is the
Klein-Gordon equation for the evolution of the field, which is
substantially different from the case of ordinary cosmologies: 
\be
\label{KG} 
\phi^{\prime\prime}+2{\cal H}\phi^{\prime}+
\frac{1}{2\omega}\left(\omega_{,\phi}\phi^{\prime\,2}-
a^{2}\frac{f_{,\phi}}{\kappa}+2a^{2}V_{,\phi}\right)= 0\, . 
\ee
As we stress in the next Section, the relevant changes with
respect to the standard picture are represented by the change in
time of the function $f$. In EQ scenarios the dynamics of the 
field possesses two distinct regimes. At low redshift, the 
behavior of the energy density coincides with the 
corresponding one in the tracking trajectories in ordinary quintessence 
models, linked to the potential exponent as $w_{\phi}=-2/(2+\alpha )$. 
At high redshift, generally much earlier than the epoch of 
structure formation, eventually the effective potential coming 
from the non-minimal interaction with gravity takes over ($R-$boost), 
and imprints a behavior $w_{\phi}=-1/3$ for the quadratic coupling 
(\ref{EQ}) \cite{baccigalupi_matarrese_etal,matarrese_baccigalupi_etal}. 

\subsection{Linear cosmological perturbations}
\label{lcp}

We will describe the linear cosmological perturbations in the real
as well as in the Fourier space. For this reason, we follow a 
notation close to the one introduced recently by Liddle and Lyth
\cite{liddle_lyth}, which allows not to explicitate the Laplace
operator eigenfunctions when working in the Fourier space,
minimizing the formal changes needed to go from the real to the
Fourier space and viceversa. See \cite{kodama_sasaki} for the
usual formulation cast in the Fourier space. \\
The general expression for the linear perturbation to the metric
(\ref{FRW}) can be written as \cite{liddle_lyth}
\be
\label{pm}
ds^2 = a^2(\tau)\{- (1 + 2\,A) d\tau^2 - B_i d\tau dx^i +
[(1 + 2\,D)\delta_{ij} + 2\,E_{ij}]dx^idx^j\},
\ee
where the function $E_{ij}$ is chosen to be traceless in order
to uniquely identify the non-diagonal spatial perturbation. \\
It is usual to further decompose the above quantities $B_i$ and
$E_{ij}$ into pure scalars ($S$), scalar-type ($S$) and
vector-type ($V$) components of vectors, scalar-type ($S$),
vector-type ($V$) and tensor-type ($T$) components of tensors,
according to their behavior with respect to a spatial coordinate
transformation:
\bea & A = & A^S ;\\
& B_i = & B_i^S + B_i^V ; \\
& E_{ij} = & E_{ij}^S + E_{ij}^V + E_{ij}^T ;\\\nonumber
\eea
where
\bea
& B_i^S = \nabla_i B, &
E_{ij}^S = \left(\nabla_i\nabla_j - \frac{1}{3}\delta_{ij}\nabla^2\right) E\ , \\
& \nabla^i B_i^V = 0, &
E_{ij}^V = \frac{1}{2}\left(\nabla_i E_j + \nabla_j E_i\right)\ ,
\ E_{i}^{i\, V}=0\ ,\\
& E_{i}^{j\, T}=0, & \nabla^i E_{ij}^T = 0, \\\nonumber \eea and
$E$, $B$ are scalar functions. Our notation slightly differs from
the one of Liddle and Lyth which is given in the Fourier space:
the quantities $E$, $E_{i}$ and $B$ we use here correspond to the
original ones divided by $k^{2}=k_{i}k^{i}$ and $k$, respectively.
In the linear theory the different types of perturbations evolve
independently of each other and can thus be treated separately. \\
An analogous decomposition can be performed for the stress-energy
tensor, whose expression up to the first perturbative order is:
\bea
& \tilde{T^0_0} = & - (\rho + \delta \rho)\ , \\
& \tilde{T^0_i} = & (\rho + p)(v_i - B_i)\ , \\
& \tilde{T^i_0} = & - (\rho + p) v^i\ , \\
& \tilde{T^i_j} = & (p + \delta p)\delta^i_j + p \Pi_j^i\ , \\\nonumber
\eea
where the fluid velocity $v_i$ and the anisotropic stress $\Pi_{ij}$
can be split as
\bea
& v_i = & v_i^S + v_i^V ; \\
& \Pi_{ij} = & \Pi_{ij}^S +\Pi_{ij}^V + \Pi_{ij}^T ,\\\nonumber
\eea
with the same properties of their metric counterparts:\\
\bea
& v_i^S = \nabla_i v\ , &
\Pi_{ij}^S = \left(\nabla_i\nabla_j - 
\frac{1}{3}\delta_{ij}\nabla^2 \right) \Pi\ ,\\
& \nabla^i v_i^V = 0\ ,&\Pi_{ij}^V = \frac{1}{2}\left(\nabla_i \Pi_j +
\nabla_j \Pi_i\right)\ ,\ \Pi_{i}^{i\, V}=0\ , \\
& \nabla^i \Pi_i = & \nabla^i \Pi_{ij}^T = 0\,, \\ \nonumber \eea
where again $\Pi$ is a scalar quantity and the same differences of
our notation with the one by Liddle and Lyth hold here for the
stress-energy tensor perturbations.

In this paper we will take into account only density (i.e.
scalar-type) perturbations. The reason is that also in the
generalized scenarios we consider here they play the dominant
role. In fact, as we stress in detail in the following, the
scalar-tensor coupling does generate a non-null anisotropic stress
already at the linear level, but that is of scalar-type only, and
therefore
does not act as a source for gravitational waves. \\
We work in the so called conformal Newtonian gauge, where
the non-diagonal perturbations to the metric are set to zero:
\be
 B \: =\:  E\:  =\:  0\,.
\ee 
Furthermore, we will rename the lapse function $A$ and the
spatial diagonal perturbation $D$ after the widely used
gauge-invariant potentials \cite{mukhanov_etal}: 
\bea & A \to
{\Psi} \,; \quad & D \to {\Phi}\,. \eea The line element used
through the rest of the paper will therefore be \be ds^2 = a^2[-
(1 + 2\, {\Psi} )d\tau^2 + (1 + 2\,{\Phi})dl^2]\,, 
\ee 
where $dl^{2}$ is the unperturbed spatial length element from
(\ref{FRW}).

We now write down the main equations driving the evolution of the
perturbed quantities defined above. For each fluid component, the
evolution of the scalar perturbed quantities can be followed
through the dynamical variables $\delta_{x} = \delta\rho_{x}
/\rho_{x}$, $v_{x}$, $\delta p_{x}$, $\Pi_{x}$, defined in terms
of the stress energy tensor as 
\be \label{deltaT} 
\delta\rho_{x}=
- \delta T^{0}_{0\, x} \  ,\ \delta p_{x}= 1/3 \:
\delta T^{i}_{i\, x} \ ,\ \nabla^i v_{x} = -\delta
T^{i}_{0\, x}/ (\rho_{x} + p_{x}) \ ,\
p_{x}\nabla_{i}\nabla^{j\ne i}\Pi_{x}=\delta T_{i}^{j\ne i}\ . 
\ee
Note that from now on we do drop the subscript $S$ meaning that we
always treat scalar cosmological perturbations, unless otherwise
specified. In the Fourier space, the equations for $\delta_{x}$
and $v_{x}$ take the form 
\be
\label{deltax} 
\delta_{x}^{\prime} - 3 {\cal{H}} w_{x} \,\delta_{x} = k^2(1 +
w_{x})v_{x} - 3(1 + w_{x}){\Phi}^{\prime} -
3{\cal{H}} w_{x}\,p_{x} \,\delta p_{x}; 
\ee
\be
\label{vx} 
 v_{x}^{\prime} + {\cal{H}}(1 - 3 w_{x})v_{x} +
\frac{w_{x}^{\prime}}{1 + w_{x}}v_{x} = - \frac{\delta
p_{x}}{\rho_{x} (1 + w_{x})} - {\Psi} + \frac{2}{3}\left(1 - \frac{3
K}{k^2}\right) \frac{w_{x}}{1 + w_{x}}\Pi_{x} \,, 
\ee 
while
$\delta p_{x}$ and $\Pi_{x}$ depend on the particular species
considered. The perturbed Klein-Gordon equation can be written in
terms of two equations formally equivalent to
(\ref{deltax},\ref{vx}) by building the conserved expression for
the perturbed energy density, pressure and anisotropic stress
perturbations \cite{perrotta_baccigalupi}. Their combination leads
to the Klein-Gordon equation at first perturbative order: 
\bea
&&\delta\phi^{\prime\prime} + \left(2\cal{H} +
\frac{\omega_{,\phi}}{\omega}\phi^{\prime}\right)\delta\phi^{\prime}
+ \left[ k^2 + \frac{1}{2}\left(
{\frac{\omega_{,\phi}}{\omega}}\right)_{,\phi} {\phi^{\prime}}^2 +
\left(\frac{- a^{2}f_{,\phi}/\kappa + 2 a^{2}V_{,\phi}}{2
\omega}\right)_{,\phi} \right] \delta\phi = \nonumber \\ &&
({\Psi}^{\prime} - 3 {\Phi}^{\prime})\phi^{\prime} + \left(2
\phi^{\prime\prime} + 4 {\cal{H}} \phi^{\prime} +
\frac{\omega_{,\phi}}{\omega}{\phi^{\prime}}^2 \right){\Psi} +
\frac{1}{2\, \omega\kappa} \frac{\partial^2 f}{\partial \phi
\,\partial R} \delta R \,. \\\nonumber 
\eea

\section{Generalized Poisson equations}
\label{gep}

In this Section we work out the equations relating the stress-energy 
tensor perturbations to the scalar metric gravitational potentials. 
The latter, together with the background cosmic geometry and expansion, 
determine entirely the lensing process. 

We start writing the generalized expression of the density
fluctuation \cite{hwang}: 
\bea && \delta\rho = \rho\cdot\delta = -
\delta T_{0}^{0} = \frac{1}{F}\Big[ \delta\rho_{\rm{fluid}} +
\frac{\omega}{a^2}\phi^{\prime}\delta\phi^{\prime} +
\frac{1}{2}\left(\frac{\omega_{,\phi}}{a^2}{\phi^{\prime}}^2 -
\frac{f_{,\phi}}{\kappa} + 2\,V_{,\phi} \right)\delta\phi -
\left(\frac{\rho + 3 p}{2} - \frac{1}{a^2}\nabla^2\right)\delta F -
\nonumber\\ 
&& \qquad  - 3 \frac{{\cal H} \delta F^{\prime}}{a^2}
+ 6 \frac{{\cal{H}}{\Psi} F^{\prime}}{a^2} - 3 \frac{{
\Phi}^{\prime} F^{\prime}}{a^2} - \frac{\omega}{a^2}{\Psi}
{\phi^{\prime}}^2 \Big]\ . 
\eea 
We can focus on two main aspects
of the generalized expression above, playing the major role into
the generalization of the Poisson equation: the contribution from
the field fluctuations $\delta\phi$ and the $1/F$ term in front of
the expression for $\delta\rho$, which acts as an effective time 
varying gravitational constant. As we shall see below, the latter 
is the relevant effect in typical Extended Quintessence models. \\
The Poisson equation relates the fluctuations in the time-time
component of the metric to the usual combination of density and
scalar-type velocity perturbations, named $\Delta$, whose
expression in Newtonian gauge is \cite{kodama_sasaki}: 
\be 
\Delta = \delta - 3\,{\cal H} w\,v\,. 
\ee 
We follow as much as possible
the notation of earlier works \cite{hu}. The 
$\delta G_{0}^{0}=\delta T_{0}^{0}$
equation can be cast in such a way that formally it coincides with
the case of ordinary cosmologies. In the Fourier space it is 
\be
\label{poisson} \frac{2}{a^2}\left(k^2 - 3\,K\right){\Phi} = 3\,\Delta
\left( H^2 + \frac{K}{a^2}\right)\,, 
\ee 
so that we can exploit
our distinction between fluid and generalized cosmology terms.
Note that the Hubble expansion rate is evaluated with respect to
the ordinary time, $H=\dot{a}/a =a^{\prime}/a^{2}={\cal H}/a$. By
using equations (\ref{generalizedrho},\ref{friedmann}), we can
write $H^2 = H^2_{\rm{fluid}} + H^2_{\rm{gc}}$, where 
\be
\label{H2fluid} 
H^2_{\rm{fluid}} =
\frac{1}{F}\frac{\rho_{\rm{fluid}}}{3} = \frac{F_0}{F}H^2_0\left[
\Omega_{0m}\left(1 + z\right)^3 + \Omega_{0r}\left(1+z\right)^{4}+ 
\left(1 - \Omega_0\right)(1 + z)^2
\right]\,, 
\ee 
and $\Omega_{0m}$, $\Omega_{0r}$ and $\Omega_0$ are the contribution 
to the present expansion rate from the matter, radiation and the total 
density respectively, 
while $F_0$ is the actual value of the gravitational 
coupling strength, and can be replaced with $1/8 \pi G$. It is important 
to note that $H^{2}_{\rm{fluid}}$ is linked to the energy density of
the fluid components, all them but the scalar field, but it
contains a most important generalization, represented by the
$F_{0}/F$ term, which plays the role of a time dependent
gravitational constant into the Friedmann equation. Moreover, 
using the relation $K/a^2 =H_0^2(\Omega_0 - 1)(1+ z)^2$, 
we included the effect of the spatial curvature into 
$H_{\rm fluid}$. The expressions 
for $H^{2}_{\rm gc}$ and $\Omega_{\rm gc}$ can be easily 
obtained by making use of the equations 
(\ref{generalizedrho},\ref{friedmann}): 
\be
\label{h2gc}
H^{2}_{\rm gc}=\rho_{\rm gc}\ \ ,
\ \ \Omega_{\rm gc}=\frac{H_{\rm gc}^{2}}{H_{0}^{2}}\,.
\ee
Starting from (\ref{poisson}), let us write down the
relation between the power spectra of ${\Phi}$, $P_{{
\Phi}}=k^{3}{\Phi}^{2}/2\pi^{2}$, and the one of $\Delta$: 
\bea
\label{gpe} && P_{{\Phi}} =
\frac{9}{4}\left(\frac{H_0}{k}\right)^4 \left(1 - \frac{3 (1 -
\Omega_0) {H_0}^2}{k^2}\right)^{- 2}\,
\cdot \nonumber \\
&&\qquad\quad \left[
\frac{F_{0}}{F}\Omega_{0m} (1 + z) + 
\frac{F_{0}}{F}\Omega_{0r} (1 + z)^{2} + 
(1 - \Omega_0) \left(\frac{F_{0}}{F} - 1\right) +
\frac{1}{(1+z)^{2}}\Omega_{\rm gc}\right]^2
P_\Delta\,.
\eea 
This is the equation which generalizes the link between
time-time metric fluctuations with density and scalar-type
velocity perturbations. The new effects arise from the scalar
field contribution, encoded in $H^{2}_{\rm gc}$, the $1/F$ term
behaving as a time dependent gravitational constant, as well as
from the fluctuations of the scalar field, contained in $\Delta$,
both in $\delta\rho$ and $v$. \\
Let us check the most relevant corrections to the quantities 
above in EQ models we take as reference 
\cite{perrotta_matarrese_etal,baccigalupi_matarrese_etal}, where 
the scalar field fluctuations $\delta\phi$ play a 
minor role \cite{perrotta_baccigalupi}, and 
the overall geometry is assumed to be flat, $K=0$. In these condition, 
the most important correction is represented by the $1/F$ term, 
effectively representing the time variation of the gravitational 
constant. The expression for $F$ in (\ref{EQ}) can be conveniently 
rewritten as 
\be
\label{EQ2}
F=\frac{1}{8\pi G}+\xi (\phi^{2}-\phi^{2}_{0})\ ,
\ee
to make explicit that at the present $F=1/8\pi G$. The observational 
constraints \cite{will,bertotti_etal} usually are expressed as bounds 
on the quantities 
\be
\frac{1}{G}\frac{dG}{dt}\simeq -\frac{1}{F}\frac{dF}{dt}=\frac{2}{\phi}
\frac{d\phi}{dt}
\ \ ,\ \ \omega_{JBD}=\frac{F}{F_{\phi}^{2}}=
\frac{F}{4\xi^{2}\phi^{2}}\,,
\ee
calculated at the present time, 
where the $\simeq$ sign above is due to the slight difference with 
the gravitational constant measured in Cavendish like experiments 
\cite{esposito-farese_polarski} and $\omega_{JBD}$ is the usual 
Jordan-Brans-Dicke parameter, which usually implies the strongest 
constraint. Typically 
\cite{perrotta_baccigalupi_etal,baccigalupi_matarrese_etal} 
the correction to the 
$1/8\pi G$ term is small, so that 
\be
\label{EQ:1overF}
\frac{1}{F}\simeq 8\pi G[1-8\pi G\xi (\phi^{2}-\phi_{0}^{2})]\ .
\ee
Moreover, in tracking trajectories with inverse power law 
potentials, the field approaches $\phi_{0}$ of the order of 
$1/\sqrt{G}$ from below, being generally 
much smaller than that in the past, when $1/F$ 
freezes to the value $8\pi G (1+8\pi G\xi\phi_{0}^{2})$. 
The magnitude of the correction is therefore 
\be
\label{EQ:correction}
8\pi G\xi\phi_{0}^{2}=\frac{1}{4\xi\omega_{JBD}}\ =
\phi_{0}\sqrt{\frac{2\pi G}{\omega_{JBD}}}\,;
\ee
note that as $\omega_{JBD}$ approaches infinity, recovering 
general relativity, the correction may still be 
relevant depending on the values of $\xi$ or $\phi_{0}$. 
The gravitational potential receives a contribution which is 
\be
\label{EQ:Phi}
\delta{\Phi}=-
8\pi G\xi(\phi^{2}-\phi_{0}^{2}){\Phi}_{{\rm fluid}+\phi}\ .
\ee
The subscript $()_{{\rm fluid}+\phi}$ represents all the terms 
coming from the fluid quantities as well as the scalar field 
ones from the minimal coupling, 
i.e. not involving $F$ explicitly: 
\be
\label{EQ:Phifluidphi}
k^{2}{\Phi}_{{\rm fluid}+\phi}=
4\pi G\Delta_{{\rm fluid}+\phi}H_{0}^{2}
[\Omega_{0m}\left(1+z\right)^{2}+
\Omega_{0r}\left(1+z\right)^{4}+\Omega_{\phi}]\ .
\ee
Note also that if the trajectory is tracking, with an almost constant 
equation of state $w_{\phi}$, the expression above reduces to 
\be
\label{EQ:Phifluidphi2}
k^{2}{\Phi}_{{\rm fluid}+\phi}=
4\pi G\Delta_{{\rm fluid}+\phi}H_{0}^{2}[
\Omega_{0m}\left(1+z\right)^{2}+
\Omega_{0r}\left(1+z\right)^{4}+
\Omega_{\phi 0}\left(1+z\right)^{3(1+w_{\phi})}]\ ;
\ee
in order to keep the notation simple, we drop such subscript in 
the following, always meaning that it is there when discussing 
approximate expressions in EQ models.  
Similarly, the gravitational potential power spectrum gets an 
extra contribution which at the first order in the correction to 
$1/8\pi G$ is 
\be
\label{EQ:PPhi}
\delta P_{\Phi}=-16\pi G\xi (\phi^{2}-\phi_{0}^{2})P_{\Phi}\ ,
\ee
where 
\be
\label{EQ:PPhifluidphi}
P_{\Phi}=\frac{9}{4}\left(\frac{H_{0}}{k}\right)^{4}
\left[
\Omega_{0m}\left(1+z\right)+
\Omega_{0r}\left(1+z\right)^{2}+
\Omega_{\phi 0}\left(1+z\right)^{1+3w_{\phi}}
\right]P_{\Delta}\,.
\ee 
In general, another important effect which arises in cosmology 
from the generalization of the underlying gravity physics 
is represented by the relation between $\Psi$ and $\Phi$. 
The difference between $f$ and R, which may arise from a 
scalar-tensor coupling as well as a non-standard dependence 
of $f$ from $R$ itself, gives origin to tidal forces exciting 
the anisotropic stress of scalar origin \cite{hwang}; in the 
Fourier space its simple form is
\begin{equation}
\label{stress}
p_{\phi}\Pi_{\phi}= \frac{k^{2}}{a^{2}}\frac{\delta F}{F}\ ,
\end{equation}
and implies a shift between the two gauge independent scalar metric
perturbations, which in our gauge takes the form
\begin{equation}
\label{PsiplusPhi} {\Psi}+{\Phi}=-\frac{a^{2}}{k^{2}}p\Pi =
-\frac{\delta F}{F}\,,
\end{equation}
where the last equality holds if the anisotropic stress is due to 
the generalization of gravity and is does not come from matter 
or radiation. This is an important
new aspect of generalized cosmologies which implies a 
change in almost all the equations describing the weak lensing
effect in cosmology, to be discussed next. For this reason, it is
convenient to give a name the ${\Psi} +{\Phi}$ combination,
valid both for the real and the Fourier space: 
\be \label{Xi}
\Xi={\Psi}+{\Phi}\,. 
\ee 
One has $\Xi=0$ in ordinary cosmology,
and $\Xi =-\delta F/F$ in the
generalized scenarios of interest here. As we already stressed, 
$\Xi$ is excited both by a scalar-tensor coupling like in EQ models, 
and a generalized dependence of the gravitational Lagrangian term 
on $R$; its expression in terms of $f$ is 
\be
\label{XiRphi}
\Xi = - \left(\frac{\partial f}{\partial R}\right)^{-1}\left[\frac{\partial^{2}f}{\partial\phi\partial R}\delta\phi
+ \frac{\partial^{2}f}{\partial R^{2}}\delta R \right]\, .
\ee
While in the first case the correction is small because of the 
smallness of the scalar field fluctuations $\delta\phi$ 
\cite{perrotta_baccigalupi,perrotta_baccigalupi_etal}, the 
contribution from the second term has not been investigated yet. 

\section{Lensing equation}

Before working out the generalized expression for the weak lensing
power spectrum it is necessary to reconsider the lensing equation
to track the new effects coming from background dynamics and
perturbations. We follow the approach by \cite{kaiser}, deriving
the photon trajectories as solutions with $ds^2 = 0 $ of the
geodesic equation for the metric (\ref{pm}); the lensing
deflection around a given direction in the sky is described
introducing new angular coordinates $\theta_x$ and $\theta_y$,
defined as \be \label{newangles} \theta_x = \theta
\cos\varphi;\quad \theta_y = \theta \sin\varphi\,, \ee where
$\theta =\sqrt{\theta_{x}^{2}+\theta_{y}^{2}}$ and $\varphi$
are the polar coordinates in the $(\theta_{x},\theta_{y})$ plane. \\
Furthermore, we perform a change of the radial coordinate:
\be
\label{chi}
\chi(r) = \frac{1}{\sqrt{K}} \arcsin \sqrt{K} r 
\ee
such that the spatial background metric takes the more readable form:
\be dl^2 = d\chi^2 + \frac{\sin^2 \sqrt{K} \chi}{K}(d\theta_x^2 +
d\theta^2_y)\,. \ee This notation is convenient because the weak
lensing hypothesis immediately reflects in the condition \be
\label{shift} \theta  \ll 1\,, \ee which allows to write the
geodesic equation at first order in the deflection angle, in
addition to the usual linear approximation for metric
perturbations. The geodesic equation is indeed \be \label{ge}
\frac{d^2 r^\alpha}{d\lambda^2} = -
g^{\alpha\beta}\left(g_{\beta\nu,\mu} -
\frac{1}{2}g_{\mu\nu,\beta}\right) \frac{dr^\mu}{d
\lambda}\frac{dr^\nu}{d\lambda}\,, \ee which for $\alpha =0$ and
$\alpha =1$ gives $d\tau = 1/a^2 d\lambda$ and $dr = 1/a^2
d\lambda$; substituting these expressions into the perturbed
equation for the angular part we get 
\be 
\label{le0}
\frac{d^2\theta_x}{d \tau^2} = -
\frac{K}{\sin^2{\sqrt{K}\chi}}\partial_{\theta_x}({\Psi} - {
\Phi}) - 2
\,\sqrt{K}\,\frac{\cos{\sqrt{K}\chi}}{\sin{\sqrt{K}\chi}} \frac{d
\theta_x}{d \tau}\, 
\ee 
for the angular coordinate $\theta_{x}$
and another one for $\theta_{y}$, formally equivalent with
$x\rightarrow y$.\\
In terms of $\Xi$, defined in (\ref{Xi}), the geodesic equation
assumes the familiar form \cite{kaiser,bartelmann_schneider} plus
the perturbation coming from the anisotropic stress ($i$ stands
now for $x$ or $y$): 
\be 
\label{le} 
\frac{d^2\theta_{i}}{d \tau^2} = 
\frac{K}{\sin^2{\sqrt{K}\chi}}\partial_{\theta_{i}}(2{\Phi} -\Xi) -
2 \,\sqrt{K}\,\frac{\cos{\sqrt{K}\chi}}{\sin{\sqrt{K}\chi}}
\frac{d \theta_{i}}{d \tau}\,. 
\ee 
The effects coming from the modified cosmological expansion are 
encoded in $\chi$, through the modified dependence of the distances 
$r$ given by (\ref{friedmann_solved},\ref{rz}) with respect to the 
redshift $z$. \\
In terms of the comoving displacement from the polar axis, $x_i =
\theta_i \sin{\sqrt{K}\chi}/\sqrt{K}$, the equation (\ref{le})
simply reads 
\be x_i^{\prime\prime} + K \,x_i = - \frac{\partial
({\Psi} - {\Phi})}{\partial x_i}=
2\partial_{i}{\Phi}-\partial_{i}\Xi\,, \ee where the first term on
the left hand side describes the tendency of two nearby rays to
converge, diverge or remain parallel according to the geometry of
the universe, while the right hand side accounts for the lensing
effect due to the metric perturbations. The general solution of
this equation is
$$
x_i + A \frac{\sin{\sqrt{K} \chi}}{\sqrt{K}} + B\,\cos{\sqrt{K}
\chi} = - \int^\chi_0 d\chi^{\prime} \partial_i
[{\Psi}(\hat{n},\chi^{\prime}) - {\Phi}(\hat{n},\chi^{\prime})]\, 
\frac{\sin{\sqrt{K}
\chi^{\prime}}}{\sqrt{K}}=
$$
\be
\label{x_i}
=\int^\chi_0 d\chi^{\prime} \partial_i [2{\Phi}(\hat{n},\chi^{\prime}) 
-
\Xi(\hat{n},\chi^{\prime})]\,
\frac{\sin{\sqrt{K} \chi^{\prime}}}{\sqrt{K}}\,,
\ee
where $A$ and $B$ are integration constants, and the position $\vec{x}$
on the light cone is completely specified by the line of sight direction
$\hat{n}$ and the generalized radial coordinate $\chi$. 

\section{Weak lensing}
\label{wl}

It is convenient to begin with the 
comoving separation between two lensed
rays, starting from the same point, one in the direction of the
polar axis and the other one in a direction $\hat{n}$, on a source
plane at distance $\chi_s$; in terms of the angular separation
$\theta_i$ in the direction $i$ one has 
\be 
x_j(\chi_s) = \left(\delta_{ij} -
\psi_{ij}\right)\frac{\sin{\sqrt{K}\chi}}{\sqrt{K}}\theta_i\,,
\ee 
where $\psi_{ij}$ is the \emph{distortion tensor}:
$$
\psi_{ij}(\hat{n},\chi) =
\frac{1}{\sqrt{K}}\int^\chi_0 d\chi^{\prime}
\frac{\sin{\sqrt{K}\chi^{\prime}}
\sin{\sqrt{K}(\chi - \chi^{\prime})}}{\sin{\sqrt{K}\chi}}
\partial_i\partial_j [{\Psi}(\hat{n},\chi') -
{\Phi}(\hat{n},\chi')]=
$$
\be
\label{distortion}
= \frac{1}{\sqrt{K}}\int^\chi_0 d\chi^{\prime}
\frac{\sin{\sqrt{K}\chi^{\prime}}
\sin{\sqrt{K}(\chi - \chi^{\prime})}}{\sin{\sqrt{K}\chi}}
\partial_i\partial_j [-2{\Phi}(\hat{n},\chi')+\Xi(\hat{n},\chi')]\,.
\ee 
Note that at first order the effect of cosmological perturbations
can be computed along the unperturbed trajectories, which corresponds
to neglect the difference in the relative deviation of two lensed and unlensed
rays inside the above integral. \\
The components of $\psi_{ij}$ are usually interpreted in terms of
the \emph{shear} $\gamma = \gamma_1 + i \gamma_2$ and of the
\emph{effective convergence} $\kappa$, respectively identified as
\bea
&& \gamma_1 = \frac{1}{2} (\psi_{11} - \psi_{22}) \,; \quad
\gamma_2 = \psi_{12}\,; \nonumber \\
&& \kappa = \frac{1}{2} (\psi_{11} + \psi_{22})\,.
\eea
Moreover, as we usually deal with lensing phenomena from a multiplicity of sources,
the distortion tensor, and thus the projected potential, are usually meant
to be integrated over the possible source distances:
\be
\psi_{ij}(\hat{n}) = \int d\chi \,g(\chi)\, \psi_{ij}(\hat{n},\chi)\,,
\ee
where $g(\chi)$ is a normalized function describing the distribution
of the relevant sources. When one considers the effect of lensing on the CMB,
the source distribution may be replaced by a delta function at the last scattering
surface. \\
Let us evaluate the correction to the distortion tensor in EQ models. 
The contribution from $\Xi$ is negligible in these scenarios, since it 
arises from the scalar field fluctuations $\delta\phi$, yielding a 
correction which is small with respect to the one coming from the 
variation of $1/F$ \cite{perrotta_baccigalupi}. Therefore, 
in flat cosmologies and at a given $\chi$, the correction to 
$\psi_{ij}(\hat{n},\chi)$ is due only to the shift in the gravitational 
potential, represented by (\ref{EQ:Phi}): 
\be
\label{EQ:psi}
\delta\psi_{ij}(\hat{n},\chi )=16\pi G\xi\phi_{0}^{2}
\int^\chi_0 d\chi^{\prime}
\frac{\chi^{\prime}(\chi - \chi^{\prime})}{\chi}
\left(\frac{\phi^{2}}{\phi_{0}^{2}}-1\right)
\partial_i\partial_j {\Phi}(\hat{n},\chi')\,.
\ee 
Note however that integrating in the $\chi$ variable, although convenient 
in order to minimize the formal corrections to $\psi_{ij}$, hides the 
effect 
of the varying gravitational constant on $\chi$ itself; in flat 
cosmologies the latter coincides with $r$ given by (\ref{rz}), which 
has to be corrected as 
\be
\label{EQ:rz}
\delta r=4\pi G \xi \phi_{0}^{2}\int_{0}^{z}\frac{dz}{H(z)}
\left(\frac{\phi^{2}}{\phi_{0}^{2}}-1\right)\ ,
\ee
as it can be easily verified since $H\propto 1/\sqrt{F}$. \\

\subsection{Generalized lensing potential}
\label{glp}

The lensing equation is often rewritten in terms of the
\emph{projected potential} $\phi$, defined through the relation
\be \psi_{ij} = \frac{K}{(\sin{\sqrt{K} \chi})^2}\, \partial_i
\,\partial_j \,\phi\,. \ee Indicating the radial coordinate
distance with $D(\chi) = \sin{\sqrt{K} \chi}/\sqrt{K}$, we get
$$
\phi(\hat{n}) = \int_0^{\chi_\infty} d\chi D(\chi)\,
[{\Psi}(\hat{n},\chi) -  {\Phi}(\hat{n},\chi)]\int^{\chi_\infty}_\chi 
d\chi^{\prime}
\frac{D(\chi^{\prime} -
\chi)}{D(\chi^{\prime})}g(\chi^{\prime})=
$$
\be \label{lp} = \int_0^{\chi_\infty} d\chi D(\chi)\,
[-2{\Phi}(\hat{n},\chi) + \Xi(\hat{n},\chi)]\int^{\chi_\infty}_\chi
d\chi^{\prime} \frac{D(\chi^{\prime} -
\chi)}{D(\chi^{\prime})}g(\chi^{\prime})\,, \ee where
$\chi_\infty$ stands for the comoving distance at infinite
redshift. By defining the integral involving the source
distribution as 
\be 
\label{sourcedistribution} 
g^{\prime}(\chi) =
{D(\chi)}\, \int^{\chi_\infty}_\chi d\,\chi^{\prime} \frac{D
(\chi^{\prime} - \chi)}{D(\chi^{\prime})} g(\chi^{\prime})\,,
\ee 
the lensing potential (\ref{lp}) takes the compact form 
\be
\label{phicompact} 
\phi(\hat{n}) = \int^{\chi_\infty}_0
d\,\chi\,g^{\prime}(\chi) \left[{\Psi}(\hat{n},\chi) - 
{\Phi}(\hat{n},\chi)\right] = \int^{\chi_\infty}_0
d\,\chi\,g^{\prime}(\chi) \left[-2{\Phi}(\hat{n},\chi) +
\Xi(\hat{n},\chi)\right]\,.
\ee 
The expression above acquires several 
new contributions in the generalized scenarios of interest here.
The modified background expansion affects the angular diameter
distance as well as the effective gravitational constant; the
perturbations get new contributions from the field fluctuations
affecting ${\Phi}$ and exciting the metric fluctuation mode 
represented by $\Xi$. \\
In EQ models, if the integration is made on the variable $\chi$, 
the main correction is due to the time variation of the effective 
gravitational constant: 
\be
\label{EQ:phicompact}
\delta\phi (\hat{n})=16\pi G\xi\phi_{0}^{2}\int^{\chi_\infty}_0
d\,\chi\,g^{\prime}(\chi) \left(\frac{\phi^{2}}{\phi_{0}^{2}}-1\right)
{\Phi}(\hat{n},\chi)\,.
\ee
In this equation, and in Eqs. 
(\ref{EQ:clpp}),(\ref{EQ:Pk}),(\ref{EQ:ctp}), 
the quantity ${\Phi}$ is meant to be  ${\Phi}_{\rm{fluid}+ \phi}$ as 
defined in Eq. (\ref{EQ:Phi}) . \\
 We need now to track these effects into the angular power
spectrum of the projected lensing potential. That is defined as
usual as 
\be \label{spectrum} C^{\phi\phi}_{l} =  \langle \,
\left| \phi_{lm}\right|^2 \,\rangle \ \ ,\ \ \phi_{lm}=  - 2 \int
d\Omega_{\hat{n}}\, \phi(\hat{n})\,Y_{lm}(\hat{n}) 
\ee 
where the
$-2$ is purely conventional in order to keep the notation
consistent with earlier works \cite{hu}. One needs now to expand
the metric fluctuations in the Fourier space with respect to the
position $\vec{x}=r\cdot \hat{n}$. The expansion functions are
just the eigenfunctions $Y_{\vec{k}}(\vec{x})$ of the Laplace
operator in curved spacetime, defined in general in curved FRW
geometry \cite{kodama_sasaki}. Their radial and angular
dependences are further expanded in ultra-spherical Bessel
functions $u_{l}$ and scalar spherical harmonics, by exploiting
the relation \be \label{bessel} Y_{\vec{k}}(\vec{x})=4\,\pi
\sum_{l,m} i^{l} u_{l}(k\,x) Y_{lm}(\hat{k})\, Y_{lm}(\hat{x})\,,
\ee where $k$ and $x$ denote the modulus of the corresponding
vectors. By using the completeness of the spherical harmonics, and
the fact that $x$ coincides with the radial distance
$D(\chi)$, the final expression for $\phi_{lm}$ is 
\be
\label{philm} \phi_{lm} = 
\sqrt{\frac{8}{\pi}}\int^{\chi_\infty}_0 d\chi \, g^{\prime}(\chi)
\int d^{3}k\left[2\Phi(\vec{k},\chi ) - \Xi (\vec{k},\chi)\right]
i^{l} u_l (k\,D(\chi)) Y_{l m}(\hat{k}) \,Y_{l m}(\hat{n})\,.
\ee 
The lensing potential angular power spectrum (\ref{spectrum})
is therefore 
\bea C^{\phi\phi}_l & = &
\frac{32}{\pi}\int^{\chi_\infty}_0 d\chi\,
g^{\prime}(\chi)\int^{\chi_\infty}_{0} d\chi^{\prime}
g^{\prime}(\chi^{\prime}) \int d^{3}k\,\int d^{3}k^{\prime}
u_l[k\,D(\chi)] u_l [k^{\prime} D(\chi^{\prime})]
Y_{lm}(\hat{k}) Y^*_{lm}(\hat{k}') \cdot\nonumber\\
&\cdot&\left[\langle
{\Phi}(\vec{k},\chi ){\Phi}(\vec{k}',\chi' )^{*} -\frac{1}{2}\Xi
(\vec{k},\chi) {\Phi}(\vec{k}',\chi' )^{*} -\frac{1}{2}\Xi
(\vec{k}',\chi')^{*}{\Phi}(\vec{k},\chi ) +\frac{1}{4}\Xi
(\vec{k},\chi)\Xi (\vec{k}',\chi')^{*} \rangle\right]\,. 
\eea
Assuming that the statistical average above eliminates the
correlation between different Fourier modes, as well as the
dependence on the direction of the wavenumbers, 
\be \label{Gauss}
\langle {\cal A}(\vec{k},\chi)\, {\cal B}(\vec{k}',\chi') \rangle
= \langle {\cal A}(k,\chi) {\cal B}(k,\chi')\rangle \delta(\vec{k} -
\vec{k}')\,, 
\ee 
where ${\cal A}$ and ${\cal B}$ represent either
${\Psi}$ or $\Xi$ and they are meant to be ensemble
averaged, one finally gets 
\bea C^{\phi\phi}_l & = &
\frac{32}{\pi}\int^{\chi_\infty}_0 d\chi\,
g^{\prime}(\chi)\int^{\chi_\infty}_{0} d\chi^{\prime}
g^{\prime}(\chi^{\prime})
\int k^{2}dk\,u_l[k\,D(\chi)] u_l [k\,D(\chi')] \cdot\nonumber\\
&\cdot&\left[\langle {\Phi}(k,\chi ){\Phi}(k,\chi' )\rangle
-\langle \Xi(k,\chi){\Phi}(k,\chi' )\rangle
+\frac{1}{4}\langle\Xi(k,\chi)\Xi(k,\chi')\rangle\right]\,. 
\label{cppx}
\eea 
It is also
useful to write down explicitly the equivalent form of the
expression above which contains the gravitational potentials only:
\bea C^{\phi\phi}_l & = & \frac{32}{\pi}\int^{\chi_\infty}_0
d\chi\, g^{\prime}(\chi)\int^{\chi_\infty}_{0} d\chi^{\prime}
g^{\prime}(\chi^{\prime})
\int k^{2}dk\,u_l[k\,D(\chi)] u_l [k\,D(\chi')] \cdot\nonumber\\
&\cdot&\left[\frac{1}{4}\langle{ \Psi}(k,\chi ){\Psi}(k,\chi' )\rangle +
\frac{1}{4}\langle{\Phi}(k,\chi ){\Phi}(k,\chi' )\rangle
-\frac{1}{2}\langle{\Psi}(k,\chi){\Phi}(k,\chi')\rangle \right]\,,
\label{cppp}
\eea
putting in evidence the correlation between ${\Psi}$ and ${\Phi}$.  \\
From this expression we can easily infer the main correction to 
the lensing potential angular power spectrum arising in EQ cosmologies, 
using (\ref{EQ:PPhi}): 
\begin{eqnarray}
\nonumber
\delta C_{l}^{\phi\phi} & = & -512G\xi\phi_{0}^{2}
\int^{\chi_\infty}_0
d\chi\, g^{\prime}(\chi)\int^{\chi_\infty}_{0} d\chi^{\prime}
g^{\prime}(\chi^{\prime}) \cdot \\   
\label{EQ:clpp}
& & \left( {\phi^2 \over \phi_{0}^{2}} -1 \right) 
\int k^{2}dk  
 \, u_l[k\,D(\chi)] u_l [k\,D(\chi')] 
\langle {\Phi}(k,\chi ){\Phi}(k,\chi' )\rangle\,.
\end{eqnarray}
Note that the numbers here conspire to yield a quite large 
factor in front of this expression, which may render the correction above 
relevant even for values of the product $G\xi\phi_{0}^{2}$ as small as 
$10^{-3}$. 

In the following we further specialize our results computing the 
generalized expression of some quantity particularly relevant for 
observations, as well as their main corrections in EQ models. 

\subsection{Convergence power spectrum}

The convergence, represented by the trace of the distortion tensor, is 
usually used as a main magnitude of the weak lensing distortion. Indeed, 
in the weak lensing hypotheses, it coincides with the shear power spectrum 
(see \cite{bartelmann_schneider}) which has been recently observed from the 
distortion induced in the shape of background galaxies in the optical band 
\cite{bacon_etal,wilson_etal,wittmann_etal,maoli_etal,vanwaerbecke_etal}. \\
The expression we need to compute is given by
\bea
\kappa_{\text{eff}} &=& \frac{1}{2}(\psi_{11} + \psi_{22})\,
\eea
which shall be averaged over the source distribution as usual. We get
\bea
\bar{\kappa}_{\text{eff}} & = & 
\frac{1}{2}\int^{\chi_\infty}_0 \,d\chi^{\prime} \,g(\chi^{\prime})\,
\int^{\chi_\infty}_{\chi}\frac{D(\chi)\,D(\chi^{\prime } - \chi)}{D(\chi^{\prime})} 
\partial^i\,\partial_i\,[{\Psi}(\hat{n},\chi) - 
{\Phi}(\hat{n},\chi)] \,= \nonumber \\
& = & \frac{1}{2} \int^{\chi_\infty}_0 \,d\chi^{\prime} \,g(\chi^{\prime})\,
\int^{\chi_\infty}_{\chi} \frac{D(\chi)\,D(\chi^{\prime } - \chi)}{D(\chi^{\prime})} 
\partial^i\,\partial_i\,[- 2{\Phi} (\hat{n},\chi) + \Xi (\hat{n},\chi)] 
\,= 
\nonumber \\
& = & \frac{1}{2}\int^{\chi_\infty}_0 \,d\chi \,g^{\prime}(\chi)  
\partial^i\,\partial_i\,[- 2{\Phi} (\hat{n},\chi) + \Xi(\hat{n},\chi)]\,.
\eea
The two-dimensional Laplacian appearing in this equation can be safely replaced with 
its three-dimensional analogue (see \cite{bartelmann_schneider} and \cite{white_hu} for a 
numerical check of this point). 
Once this substitution has been made, we can expand the generalized 
gravitational potential in 
Fourier harmonics transforming with respect to the spatial point $\hat{n}\cdot\chi$, 
and transform the Laplacian in a multiplication by $(- k^2)$\,: 
\bea
\bar{\kappa}_{\text{eff}} & = & \frac{1}{2 (2\,\pi)^{3/2}}
\int^{\chi_\infty}_0 \,d\chi \,g^{\prime}(\chi) \int d^3k   
\:k^2\:[2{\Phi}(\vec{k},\chi)-\Xi(\vec{k},\chi)]\,
Y_{\vec{k}}(\vec{x})\,.
\eea
Comparing this expression with the ones for the lensing potential power spectrum 
(\ref{cppx},\ref{cppp}) we can immediately infer the result:
\bea
P_{\bar{\kappa}}(l) & = & 
\frac{8}{\pi}\int_{0}^{\chi_\infty} d\chi\, g^{\prime}(\chi)
\int_{0}^{\chi_\infty} d\chi^{\prime} g^{\prime}(\chi^{\prime})  
\int \,dk\,k^8\, u_l(k\,D(\chi)) u_l (k \,D(\chi^{\prime})) \cdot\nonumber\\
&\cdot&\left[\langle{\Phi}(k,\chi ){\Phi}(k,\chi' )\rangle
-\langle \Xi(k,\chi){\Phi}(k,\chi' )\rangle
+\frac{1}{4}\langle \Xi(k,\chi)\Xi(k,\chi')\rangle \right]\,=\nonumber\\
&=&\frac{8}{\pi}\int_{0}^{\chi_\infty} d\chi\, g^{\prime}(\chi)
\int_{0}^{\chi_\infty} d\chi^{\prime} g^{\prime}(\chi^{\prime})  
\int \,dk\,k^8\, u_l(k\,D(\chi)) u_l (k \,D(\chi^{\prime})) \cdot\nonumber\\
&\cdot&\left[\frac{1}{4}\langle {\Psi}(k,\chi ){\Psi}(k,\chi' ) \rangle +
\frac{1}{4}\langle {\Phi}(k,\chi ){\Phi}(k,\chi' )\rangle
-\frac{1}{2}\langle {\Psi}(k,\chi){\Phi}(k,\chi')\rangle \right]\,.
\label{pconvergence}
\eea
In EQ cosmologies, using again (\ref{EQ:PPhi}), one finds
\begin{eqnarray}
\nonumber
\delta P _{\bar{\kappa}}(l)& =& -128G\xi\phi_{0}^{2}
\int_{0}^{\chi_\infty} d\chi\, g^{\prime}(\chi)
\int_{0}^{\chi_\infty} d\chi^{\prime} g^{\prime}(\chi^{\prime})
\left(\frac{\phi^{2}}{\phi_{0}^{2}}-1\right) \cdot \\ 
\label{EQ:Pk}
& & \int \,dk\,k^8\, u_l(k\,D(\chi)) u_l (k \,D(\chi^{\prime}))
\langle {\Phi}(k,\chi ){\Phi}(k,\chi')\rangle \,,
\end{eqnarray}
where the correction to $\chi$ must be taken into account following (\ref{EQ:rz}) 
if the integration is made on the redshift. 

\subsection{Lensing in the CMB signal}

The lensing potential correlates significantly with secondary anisotropies of the CMB, 
because it arises at the epoch of structure formation; here we generalize the lensing 
cross-correlation with the Integrated Sachs-Wolfe effect (ISW, see \cite{hu} for a 
comparison with the case of ordinary cosmologies). \\
The latter can be represented in terms of temperature fluctuations as
\bea
\Theta^{ISW}(\hat{n}) &=& 
-\,\int_{0}^{\infty} d\chi [\dot{\Phi}(\hat{n},\chi ) 
-\dot{\Psi}(\hat{n},\chi )] \,=\, 
-\,\int_{0}^{\infty} d\chi \,
[2\dot{\Phi}(\hat{n},\chi)-\dot{\Xi}(\hat{n},\chi)] \,.
\eea
Note that, in order to avoid confusion with the integration variable 
$\chi'$, in this paragraph only we will denote with an overdot the 
derivative with respect to conformal time. \\
Again using the expressions for the lensing potential power 
spectra (\ref{cppx},\ref{cppp}), and making use of the statistical 
independence of different Fourier modes, we are able to write immediately 
the cross-correlated spectrum:
\bea
C^{\Theta\phi} &=& \frac{8}{\pi} 
\int_{0}^{\chi_\infty} d\chi\, g^{\prime}(\chi)\int_{0}^{\chi_\infty} 
d\chi^{\prime} \int k^2\,dk\, u_l(k\,D(\chi)) u_l (k \,D(\chi^{\prime}))\cdot\nonumber\\ 
&\cdot&\left[\langle {\Phi}(k,\chi )\dot{{\Phi} }(k,\chi' )\rangle
-\frac{1}{2}\langle \Xi(k,\chi)\dot{{\Phi}}(k,\chi' )\rangle
-\frac{1}{2}\langle \dot{\Xi}(k,\chi'){\Phi}(k,\chi)\rangle
+\frac{1}{4}\langle \Xi(k,\chi)\dot{\Xi}(k,\chi')\rangle \right]\,=\nonumber\\
&=&\frac{2}{\pi}\int_{0}^{\chi_\infty} d\chi\, g^{\prime}(\chi)\int_{0}^{\chi_\infty} 
d\chi^{\prime} \int k^2\,dk\, u_l(k\,D(\chi)) u_l (k \,D(\chi^{\prime}))\cdot\nonumber\\ 
&\cdot&\left[\langle{\Psi}(k,\chi )\dot{{\Psi}}(k,\chi' )\rangle +
\langle {\Phi}(k,\chi )\dot{\Phi}(k,\chi' )\rangle
-\langle {\Psi} (k,\chi)\dot{\Phi}(k,\chi')\rangle
-\langle \dot{{\Psi}}(k,\chi'){\Phi}(k,\chi)\rangle
\right]\,.
\label{ctp}
\eea 
The main correction in EQ cosmologies is obtained again by using (\ref{EQ:PPhi}). 
Interestingly, the time derivative reintroduces a term proportional to 
$|{\Phi}|^{2}$: 
$$
\delta C^{\Theta\phi}=-128G\xi\phi_{0}^{2}
\int_{0}^{\chi_\infty} d\chi\, g^{\prime}(\chi)\int_{0}^{\chi_\infty} 
d\chi^{\prime} \left(\frac{\phi^{2}}{\phi_{0}^{2}}-1\right)
\int k^2\,dk\, u_l(k\,D(\chi)) u_l (k \,D(\chi^{\prime}))
\langle {\Phi}(k,\chi )\dot{{\Phi}}(k,\chi')\rangle -
$$
\be
\label{EQ:ctp}
-256 G\xi\phi_{0}^{2}
\int_{0}^{\chi_\infty} d\chi\, g^{\prime}(\chi)\int_{0}^{\chi_\infty} 
d\chi^{\prime}\frac{\phi\dot{\phi}}{\phi_{0}^{2}}
\int k^2\,dk\, u_l(k\,D(\chi)) u_l (k \,D(\chi^{\prime}))
\langle {\Phi}(k,\chi ){\Phi}(k,\chi')\rangle \,.
\ee
These expressions can be further simplified noticing that for lensing on the CMB signal the source 
distribution is well represented by a delta function at the last scattering (LS) surface; thus the 
averaging function $g^{\prime}(\chi)$ can be written as
\bea
g^{\prime}(\chi)&=& {D(\chi)}\, 
\int^{\chi_\infty}_{\chi} d\,\chi^{\prime} 
\frac{D_(\chi^{\prime} - \chi)}{D(\chi^{\prime})} \,
\delta(\chi^{\prime} - \chi_{\rm{LS}})\,= 
\frac{D(\chi)\,D(\chi_{\rm{LS}} - \chi)}{D(\chi_{\rm{LS}})}\,.
\eea

\section{Conclusions}

The weak lensing in cosmology is one of the most important tools 
to investigate the structure formation process, and in particular 
the mechanics of the dark cosmological component, which represents 
almost the 95\% of the cosmic budget according to the recent 
measurements. Since the onset of cosmic acceleration occurs at the 
same epoch of structure formation, the weak lensing is most promising 
in order to gain insight into the nature of the dark energy. 
The candidates which have been proposed for explaining the dark 
energy are suitably described in a cosmological context which is 
generalized with respect to the ordinary one, admitting dark matter-energy 
couplings as well as generalized theories of gravity. \\
In particular for the latter class of theories, a systematic treatment of the 
weak lensing process lacks in literature and this work aims at filling this 
gap. We considered a Lagrangian where the gravitational sector is made of a 
function which depends arbitrarily on the Ricci scalar as well as on 
a scalar field; the most general scalar-tensor theory of gravity, as well as 
the most general dependence on the Ricci scalar without a scalar field, 
can be described in full generality in this framework. \\
We studied the generalized 
Poisson equations linking the fluctuating components to the two gauge invariant 
generalized gravitational potentials representing the metric fluctuations which 
cause the weak lensing process itself. This allowed to fix the contributions from 
the modified background expansion as well as the fluctuations, both in the Ricci 
scalar and in the scalar field responsible for the scalar-tensor coupling. 
We show that both of them are responsible for an anisotropic stress of scalar 
origin, causing the gravitational potentials to be different already at the 
linear level. 
We studied in particular the modifications induced by the time variation of the 
effective gravitational constant, which are most relevant in non-minimally coupled 
models in which the gravitational Lagrangian sector is a product of a function 
depending on a scalar field and on the Ricci scalar; we focus in particular 
on the Extended Quintessence (EQ) scenarios, where the scalar field playing the 
role of the dark energy and responsible 
for cosmic acceleration today possesses a quadratic coupling with 
the Ricci scalar. \\
Starting from the equation describing the geodesic deviation, we derived 
the generalized expressions for distortion tensor and projected lensing 
potential, tracking the effects due to the time variation of the effective 
gravitational constant and the contribution of the anisotropic stress; 
in particular, we show how the latter yields a correction proportional to 
the correlation between the gravitational potentials. \\
Finally, we specialized our results to the description of two quantities 
which are most relevant for observations, i.e. the lensing convergence power 
spectrum as well as the correlation between the lensing potential and the 
Integrated Sachs-Wolfe (ISW) effect affecting the total intensity and polarization 
anisotropies in the cosmic microwave background radiation. \\
By considering again the particular case of EQ cosmologies, 
we worked out approximate expressions for the corrections induced 
by the time variation of the effective gravitational constant on the 
lensing potential, lensing-lensing correlation angular power spectrum, 
convergence angular power spectrum as well as lensing-ISW correlation. 
We showed that the order of magnitude of these effects is of the order 
of $8\pi G\xi\phi_{0}^{2}$, where $\xi$ is the coupling constant and $\phi_{0}$ 
is the present value of the dark energy field. 
It may be noted how such correction may be relevant even if the underlying 
theory is close to general relativity, i.e. if the Jordan-Brans-Dicke parameter 
$\omega_{JBD}=1/32\pi G\xi^{2}\phi_{0}^{2}$ is large, depending on the relative balance 
between $\xi$ and $\phi_{0}$. 

Despite of these interesting indications in the particular case of EQ cosmologies, 
the formulas we developed here have great generality, allowing a direct 
interpretation of the modern weak lensing observations in the context of cosmologies 
with generalized theories of gravity. 

\acknowledgements
V.A. warmly thanks Martin White and George Smoot for useful hints and 
for hosting her at the Lawrence Berkeley National Laboratories, where this 
work was initiated. 
We are thankful to Sabino Matarrese for many useful discussions and to Eric Linder for several precious comments. 

\bibliographystyle{aa}

\begin{thebibliography}{99}
\bibitem{riess_etal} A. G. Riess et al., \apj {\bf 116}, 1009 (1998)
\bibitem{perlmutter_etal}
S. Perlmutter et al., \apj {\bf 517}, 565 (1999)
\bibitem{bennett_etal} 
C. L. Bennett \etal, \apjs {\bf 148}, 97 (2003) 
\bibitem{dodelson_etal} S. Dodelson et al., \apj {\bf 572}, 140 (2002)
\bibitem{percival_etal} W. J. Percival et al., \mnras {\bf 337}, 1068 
(2002) 
\bibitem{freedman_etal} W. L. Freedman et al., \apj {\bf 553}, 47 (2001) 
\bibitem{sahni_starobinski} V. Sahni and A. Starobinski \ijmpd {\bf 9}, 
373 (2000)
\bibitem{padmanabhan} T. Padmanabhan, \pr {\bf 380}, 235 (2003) 
\bibitem{peebles_ratra} P. J. E. Peebles and B. Ratra, \rmp {\bf 75}, 
599 (2003)
\bibitem{wetterich} C. Wetterich, \np B {\bf 302}, 668 (1988) 
\bibitem{ratra_peebles} B. Ratra and P. J. E. Peebles, \prd {\bf 37}, 
3406 (1988) 
\bibitem{coble_etal} 
K. Coble, S. Dodelson and J. A. Frieman, \prd {\bf 55}, 1851 (1997) 
\bibitem{ferreira_joyce} P. G. Ferreira and M. Joyce, \prd {\bf 58}, 
023503 (1998) 
\bibitem{masiero_etal} A. Masiero, M. Pietroni and F. Rosati, \prd 
{\bf 61}, 023504 (2000) 
\bibitem{brax_martin} P. Brax and J. Martin, \prd {\bf 61}, 103502 
(2000) 
\bibitem{liddle_scherrer} A. R. Liddle and R. J. Scherrer, \prd {\bf 
59}, 023509 (1999) 
\bibitem{steinhardt_etal} P. J. Steinhardt, L. Wang and I. Zlatev, \prd 59, 123504 
(1999) 
\bibitem{doran_etal}
M. Doran, M. Lilley, J. Schwindt and C. Wetterich, 
Astrophys. J. {\bf 559}, 501 (2001) 
\bibitem{corasaniti_copeland} P. S. Corasaniti and E. J. Copeland, Phys. 
Rev. D {\bf 65}, 043004 (2002)
\bibitem{balbi_etal} A. Balbi, C. Baccigalupi, S. Matarrese, F. Perrotta 
and N. Vittorio, \apjl {\bf 547}, L89 (2001)
\bibitem{baccigalupi_balbi_etal} C. Baccigalupi, 
A. Balbi, S. Matarrese, F. Perrotta and N. Vittorio, Phys. Rev. D 
{\bf 65}, 063520 (2002)
\bibitem{caldwell_etal}
R.R. Caldwell, M. Doran, C.M. Muller, G. Schafer and 
C. Wetterich, Astrophys. J. Lett. {\bf 591}, L75 (2003) 
\bibitem{spergel_etal} D. N. Spergel \etal, \apjs {\bf 148}, 97 (2003) 
\bibitem{melchiorri_etal} A. Melchiorri, L. Mersini, C. J. Odman and M. 
Trodden, \prd {\bf 68} 043509 (2003)
\bibitem{wang_tegmark} Y. Wang and M. Tegmark, astro-ph/0403292 (2004)
\bibitem{carroll} S. M. Carroll, \prl {\bf 81}, 3067 (1998) 
\bibitem{sahni} V. Sahni and S. Habib, \prl {\bf 81}, 1766 (1998)
\bibitem{chiba} T. Chiba, \prd {\bf 60}, 083508 (1999) 
\bibitem{uzan} J. P. Uzan, \prd {\bf 59}, 123510 (1999) 
\bibitem{bartolo_pietroni} N. Bartolo and M. Pietroni, \prd {\bf 61}, 
023518 (2000) 
\bibitem{perrotta_baccigalupi_etal} F. Perrotta, C. Baccigalupi and
S. Matarrese, \prd {\bf 61}, 023507 (2000)
\bibitem{faraoni} V. Faraoni, \prd {\bf 62}, 023504 (2000) 
\bibitem{baccigalupi_matarrese_etal}
C. Baccigalupi, S. Matarrese and F. Perrotta, \prd {\bf 62}, 123510 
(2000) 
\bibitem{esposito-farese_polarski} G. Esposito-Farese and
D. Polarski, \prd {\bf 63}, 063504 (2001)
\bibitem{riazuelo_uzan}
A. Riazuelo and J. P. Uzan, \prd {\bf 66}, 023525 (2002) 
\bibitem{torres} D. F. Torres, \prd {\bf 66}, 043522 (2002)
\bibitem{perrotta_baccigalupi} 
F. Perrotta and C. Baccigalupi, \prd {\bf 59}, 123508 (2002) 
\bibitem{perrotta_matarrese_etal} F. Perrotta, S. Matarrese, 
M. Pietroni and  C. Schimd, \prd {\bf 69}, 084004 (2004)
\bibitem{linder} E. Linder, astro-ph/0402503 (2004)
\bibitem{matarrese_baccigalupi_etal} S. Matarrese, C. Baccigalupi, 
F. Perrotta, astro-ph/0403480
\bibitem{amendola_2000} L. Amendola, \prd {\bf 62}, 043511 (2000) 
\bibitem{amendola_etal} L. Amendola, C. Quercellini, D. Tocchini-Valentini 
and A. Pasqui, \apjl {\bf 583} L53 (2003) 
\bibitem{matarrese_pietroni_etal} 
S. Matarrese, M. Pietroni and C. Schimd, \jcap {\bf 08}, 005 (2003) 
\bibitem{amendola_2003} L. Amendola, astro-ph/0311175 (2003)
\bibitem{armendariz-picon_etal} C. Armendariz-Picon, V. Mukhanov and P. J. Steinhardt,
\prd {\bf 63}, 103510 (2001) 
\bibitem{caldwell} R. R. Caldwell, \plb {\bf 545}, 23 (2002) 
\bibitem{malquarti_etal} M. Malquarti, E. J. Copeland and A. R. Liddle, 
\prd {\bf 68}, 023512 (2003) 
\bibitem{fujii_maeda} Y. Fujii, K. Maeda, {\it The Scalar-Tensor Theory of 
Gravitation}, Cambridge University Press (2003) 
\bibitem{hwang} J.Hwang, \apj {\bf 375}, 443 (1991) 
\bibitem{bartelmann_schneider} M. Bartelmann and P. Schneider, \pr 
{\bf 340}, 291 (2001)
\bibitem{bacon_etal} D. Bacon, A. Refregier and R. Ellis, \mnras {\bf 
318}, 625 (2000) 
\bibitem{wilson_etal} G. Wilson, Kaiser N. and 
G. A. Luppino, \apj {\bf 556}, 601 (2001) 
\bibitem{wittmann_etal} D. M. Wittmann, 
J. A. Tyson, D. Kirkman, I. Dell'Antonio and G. Bernstein, \n {\bf 405}, 
143 (2000) 
\bibitem{maoli_etal} R. Maoli et al., \aa {\bf 368}, 766 (2001) 
\bibitem{vanwaerbecke_etal} L. van Waerbecke et al., \aa {\bf 374}, 757 
(2001) 
\bibitem{refregier} A. Refregier \araa {\bf 41}, 645 (2003) 
\bibitem{huterer} D. Huterer, \prd {\bf 65}, 063001 (2002) 
\bibitem{hu_2002} W. Hu, \prd {\bf 66}, 083515 (2002) 
\bibitem{bartelmann_etal} M. Bartelmann, 
F. Perrotta and C. Baccigalupi, \aa {\bf 396}, 21 (2002) 
\bibitem{bernardeau} F. Bernardeau, \rpp {\bf 66}, 691 (2003)
\bibitem{weinberg_kamionkowski} N. N. Weinberg and 
M. Kamionkowski, \mnras {\bf 341}, 251 (2003) 
\bibitem{bartelmann_meneghetti_etal} M. Bartelmann, M. Meneghetti, 
F. Perrotta, C. Baccigalupi and L. Moscardini, \aa 
{\bf 409}, 449 (2003) 
\bibitem{jain_taylor} B. Jain and  A. Taylor \prl {\bf 91}, 141302 
(2003)
\bibitem{giovi_etal} F. Giovi, C. Baccigalupi and F. Perrotta \prd {\bf 
68}, 123002 (2003)
\bibitem{klypin_etal} A. Klypin, A. Maccio', R. Mainini and S. 
Bonometto, \apj {\bf 599}, 31
\bibitem{dolag_etal} K. Dolag, M. Bartelmann, F. Perrotta, C. Baccigalupi, L. 
Moscardini, M. Meneghetti and G. Tormen, \aa {\bf 416}, 853 (2004)
\bibitem{maccio_etal} A. Maccio', C. Quercellini, R. Mainini, L. 
Amendola and S. Bonometto, astro-ph/0309671 (2003)
\bibitem{maccio} A. Maccio', astro-ph/0402657 (2004)
\bibitem{hu} W. Hu, \prd {\bf 62}, 043007 (2000). 
\bibitem{liddle_lyth} A. R. Liddle and D. H. Lyth, 
{\it Cosmological Inflation and Large-Scale Structure}, Cambridge University Press, 2000 
\bibitem{kodama_sasaki} H. Kodama and M. Sasaki, \ptps {\bf 78}, 1 
(1984)
\bibitem{mukhanov_etal} V. F. Mukhanov, H. A. Feldman, and R. H. Brandeberger, 
\pr {\bf 215}, 203 (1992)
\bibitem{will} C. M. Will, \lrr {\bf 4}, 4 (2001)
\bibitem{bertotti_etal} B. Bertotti, L. Iess and  P. Tortora, \n {\bf 
425}, 374 (2003)
\bibitem{kaiser} N. Kaiser \apj {\bf 498}, 26 (1998)
\bibitem{white_hu} M. White and W. Hu, \apj {\bf 537}, 1 (2000) 

\end{thebibliography}

\end{document}